# Mathematical Modeling of Electrolyte Flow Dynamic Patterns and Volumetric Flow Penetrations in the Flow Channel over Porous Electrode Layered System in Vanadium Flow Battery with Serpentine Flow Field Design

Xinyou Ke[a,b,e,*], Joseph M. Prahl[a], J. Iwan D. Alexander[a,c], and Robert F. Savinell[d,e]

[a]Department of Mechanical and Aerospace Engineering, Case Western Reserve University, Cleveland, Ohio 44106, United States

[b]John A. Paulson School of Engineering and Applied Science, Harvard University, Cambridge, Massachusetts 02318, United States

[c]School of Engineering, University of Alabama at Birmingham, Birmingham, Alabama 35294, United States

[d]Department of Chemical Engineering, Case Western Reserve University, Cleveland, Ohio 44106, United States

[e]Electrochemical Engineering and Energy Laboratory, Case Western Reserve University, Cleveland, Ohio 44106, United States

[*]Corresponding author: xinyouke@gmail.com

**Abstract**

In this work, a two-dimensional mathematical model is developed to study the flow patterns and volumetric flow penetrations in the flow channel over the porous electrode layered system in vanadium flow battery with serpentine flow field design. The flow distributions at the interface between the flow channel and porous electrode are examined. It is found that the non-linear pressure distributions can distinguish the interface flow distributions under the ideal plug flow and ideal parabolic flow inlet boundary conditions. However, the volumetric flow penetration within the porous electrode beneath the flow channel through the integration of interface flow velocity reveals that this value is identical under both ideal plug flow and ideal parabolic flow inlet boundary conditions. The volumetric flow penetrations under the advection effects of flow channel and landing/rib are estimated. The maximum current density achieved in the flow battery can be predicted based on the 100% amount of electrolyte flow reactant consumption through the porous electrode beneath both flow channel and landing channel. The corresponding theoretical maximum current densities achieved in vanadium flow battery with one and three layers of SGL 10AA carbon paper electrode have reasonable agreement with experimental results under a proper permeability.

**Nomenclature**

| | |
|---|---|
| *A* | area integration |
| *BC* | boundary dondition |
| *c* | concentration (mol cm$^{-3}$) |
| *k* | permeability of the porous electrode (cm$^2$) |
| *L* | length (cm) |
| *P* | pressure (Pa) |
| *<P>* | average pressure (Pa) |
| *Q* | volumetric flow rate (ml min$^{-1}$ or cm$^3$ s$^{-1}$) |
| *r* | distance between adjacent flow channels or length of landing channel/rib (cm) |
| *t* | thickness (cm) |
| *u* | X direction velocity (cm s$^{-1}$) |
| *<u>* | average X direction velocity (cm s$^{-1}$) |
| *v* | Y direction velocity (cm s$^{-1}$) |
| *<v>* | average Y direction velocity (cm s$^{-1}$) |
| *w* | width (cm) |
| *X* | X direction |
| *Y* | Y direction |

Greek symbols

| | |
|---|---|
| *ε* | porosity |
| *μ* | dynamic viscosity (Pa•s) |
| *υ* | kinematic viscosity (cm$^2$ s$^{-1}$) |
| $\rho$ | density of electrolyte fluid (g cm$^{-3}$) |
| $\Sigma$ | interface |

Subscripts

| | |
|---|---|
| *avg* | average value |
| *cf* | between current collector and flow channel |
| *e* | entrance |
| *f* | flow domain |
| *fc* | flow channel |
| *fp* | between flow channel and porous electrode |
| *in* | inlet |
| *i* | number (1, 2, 3, 4, 5 and 6) |
| *lc* | landing channel |
| *p* | porous domain |
| *pm* | between porous electrode and ion selective membrane |

Dimensionless number

| | |
|---|---|
| $P_f^*$ | $P_f\,(\rho u_{in}^2)^{-1}$ |
| $<P_p>^*$ | $<P_p>\,(\rho u_{in}^2)^{-1}$ |
| $u_f^*$ | $u_f\,u_{in}^{-1}$ |
| $<u_p>^*$ | $<u_p>\,u_{in}^{-1}$ |
| $v_f^*$ | $v_f\,u_{in}^{-1}$ |
| $<v_p>^*$ | $<v_p>\,u_{in}^{-1}$ |
| $X^*$ | $X\,L^{-1}$ |
| $Y^*$ | $Y\,(t_f+t_p)^{-1}$ |
| $Re$ | Reynolds number ($\rho u_{in} t_f/\mu$) |

## 1. Introduction

Redox flow batteries (RFBs) technologies have been gained unprecedented attention for medium and large-scale energy storage applications [1-3]. In conjunction with electricity generation from intermittent renewable energy sources (e.g. wind, tide and solar energy), RFBs systems are demonstrated to be alternative tools within enabling improved stability of national grid [4]. Until now, several types of RFBs have been developed, typically such as vanadium [5,6], iron-chromium [7,8], all iron aqueous/all iron slurry [9], semi-solid lithium [10], etc. Designs of new electrolytes and flow cell architectures attract researchers in the field of flow batteries. The studies on working mechanism of flow batteries are still going on. In this effort, we will study the transport mechanism in a typical all vanadium redox flow battery (VRFB) [4,11-13] and one advantage for the VRFB is that species can be reversibly consumed in the electrolyte reservoir [5]. Experiments and mathematical modeling are two approaches to study the fundamental insights. Compared with experimental studies, the capital costs and labor effort can be reduced while crucial understanding of the transport phenomena also can be achieved through the computational modeling [14,15]. Earlier work on mathematical modeling of redox batteries was reported by Newman et al. [16,17] and they proposed a one-dimensional theoretical model to study current distributions and non-uniform kinetic reactions though the porous electrode. Shah et al. [18] developed a two-dimensional vanadium flow battery without flow channel model to study the transport physics including convection, diffusion and migration in the electrode and membrane. Subsequently, You et al. [19] studied the parameter effects on the distributions of local reactant concentration, over-potential and transfer current density based on the previous reported two-dimensional flow battery model without flow field [18]. A two-dimensional transient vanadium flow battery without flow field model was developed by Knehr et al. [15]. The species crossover at the interface boundary between the electrode and membrane were studied. The worked done by Newman et al. [16,17], Shah et al. [18], You et al. [19] and Knehr et al. [15] were on the flow batteries without flow fields through the felts (e.g. carbon and graphite felts). However, Aaron et al. [20] first reported a vanadium flow cell stack configuration with serpentine flow field over carbon paper electrode. The electrochemical performance (e.g. limiting current density and peak power density) was improved. The thickness of carbon paper electrode (~0.04 cm) [20] used in the flow battery with flow field is much thinner than the graphite felt or carbon felt (typical ~0.3 cm) [18,19] used in the one without flow field. Here, the flow channels, such as serpentine and interdigitated flow channels were evolved from proton exchange membrane fuel cell (PEMFC) design [20,21]. Up to date, few theoretical studies on details of RFBs with flow fields have been reported. Ke et al. [22] developed a macroscopic model of RFBs with a single passage of the serpentine flow channel. Both numerical and analytical flow distributions in the flow channel and porous electrode were studied [23,24]. The first mathematical model for predicting the maximum current density achieved in the flow battery with flow fields was proposed. This maximum current density model is based on the consumption of total flow ion reactants through the porous electrode in the RFBs with flow field designs. The thinner carbon fiber paper electrodes (typical 10AA SGL and Toray paper) used in the RFBs with flow fields (e.g., serpentine and interdigitated flow channels) can reduce ohmic losses compared with the conventional RFBs without flow fields. The electrochemical performance of flow cell with flow field design is improved. There are two porous electrode configurations of RFBs: (1) the classic RFBs without flow field [14-19] as shown in Fig. 1 (a), the electrolyte flow through a thick electrode; (2) the RFBs with flow field [20-26] as shown in Fig. 1 (b), the electrolyte flows through the flow channel and by-pass into a thinner electrode. The fundamental studies on the flow details in the flow batteries with flow field designs are quite few. The deep mechanism of electrolyte flow reactant penetration into the porous electrode from the flow channel is not well understood. There are two typical experimental approaches to characterize the electrolyte flow dispersion: visualization [27] and residence time distribution (RTD) [28] techniques. The visualization technique can capture the transport process (e.g. streamline) of electrolyte flow and RTD can calculate the time that particles go through the reactors based on the probability function. Although both visualization and RTD approaches are always prior to multiphysics modeling on characterization of flow dispersion, it seems that they have less capabilities of capturing how much amount of electrolyte flow penetration through the porous electrode. In this study, the amount of electrolyte flow penetration from the flow channel into the porous electrode is concerned. The flow penetration and possible maximum current density achieved based on 100% utilization of electrolyte through the porous electrode is also correlated. Thus, the multiphysics modeling approach is preferred to visualization and RTD techniques in this study on electrolyte flow penetration.

The motivation of this work on vanadium flow battery with flow field design over carbon paper electrodes compared with classic one without flow field design through the carbon felts is based on two possible merits: (1) the forced convection through the flow fields can enhance the mass transport through the electrode layer and (2) thinner carbon paper electrode (hundred microns) has lower ohmic loss in contrast to the thicker carbon felts (several millimeters). In this article, we study the flow patterns in the flow channel and porous electrode to explain the role of the convection consequently mass transfer in the flow batteries with flow field design as shown in Fig. 1 (b). The flow penetration or volumetric flow rate within the porous electrode is studied under both ideal plug flow inlet and ideal parabolic flow inlet boundary conditions. The maximum current density achieved in the flow cell can be estimated by the function of calculated volumetric flow rate within the porous electrode, number of electron transferred, Faraday constant, ion concentration and contact area between the flow channel and porous electrode. This mathematical model should contribute a certain guidance on performance optimization of flow batteries with flow field designs.

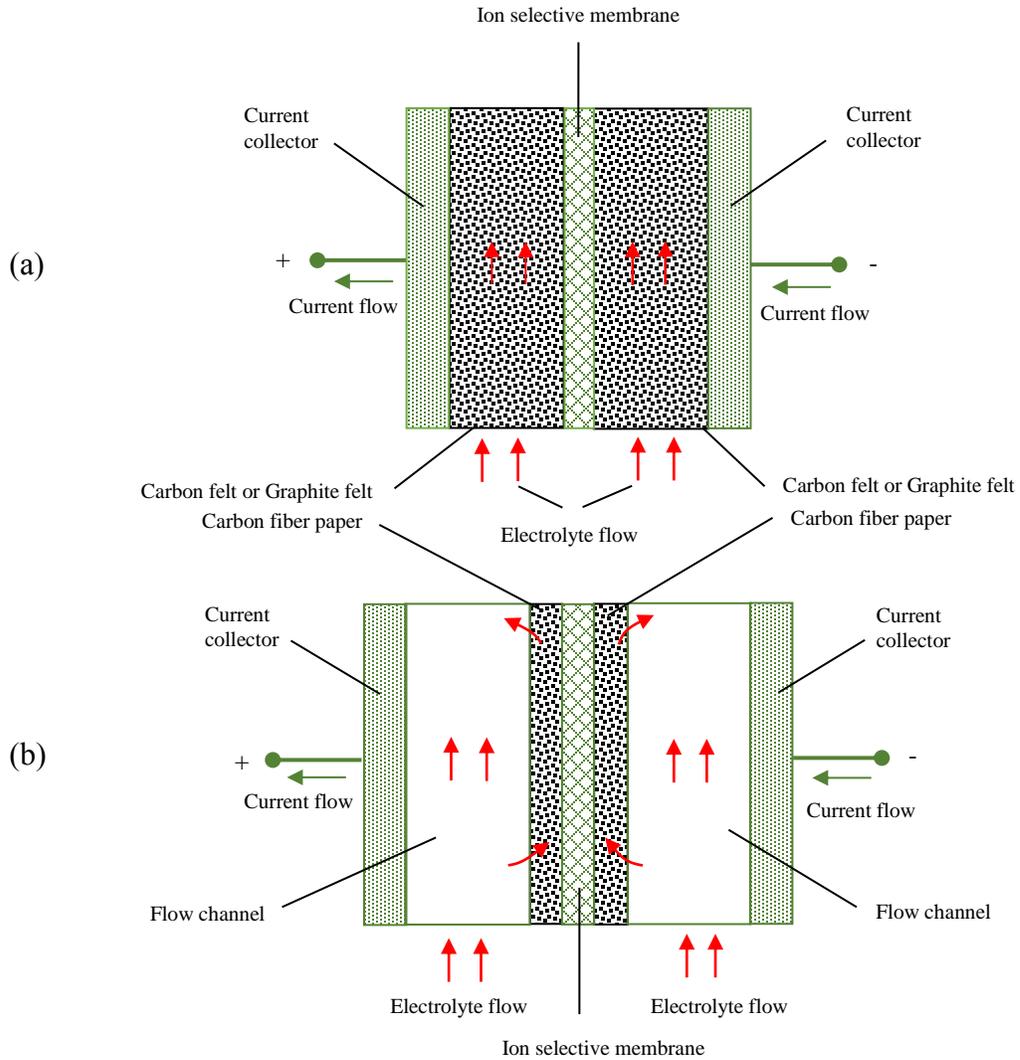

**Fig. 1.** Two-dimensional configurations of RFBs without and with flow fields: (a) classic flow cell without flow field (not drawn to scale) and (b) flow cell with flow field (not drawn to scale)

## 2. Two-dimensional Flow Cell Model

In this analysis, the simplified two-dimensional model of RFBs with flow field (see Fig. 1 (b)) is shown in Fig. 2 (a) (not drawn to scale) for Fig. 2 (b) (not drawn to scale, see (b. 1)). The actual three-dimensional model of a serpentine flow channel over the porous electrode as shown in Fig. 2 (b) (not drawn to scale, see (b. 2)) is simplified assumed as several periodic components of straight flow channels (see (b. 1)) and landing channels (see (b. 3)). The volumetric flow penetrations within the porous electrode under the advection effects from both flow channel and landing channel are considered. In this paper, the volumetric flow penetration under the rib advection is estimated through the calculated volumetric flow penetration under the flow channel advection. In the flow channel, the corresponding non-dimensional flow motions along the X and Y directions are governed by Navier-Stokes equation as shown in Eqs. (1) and (2). The flow physics in the porous electrode are described by Brinkman-Darcy model [22] with non-dimensional forms as shown in Eqs. (3) and (4) along the X and Y directions, respectively. The non-dimensional terms are defined in the Nomenclature section. The dimensionless equations that describe the flow physics in the flow channel and porous electrode are given

$$Re \frac{(t_f+t_p)^2}{t_f \cdot L} u_f^* \frac{\partial u_f^*}{\partial X^*} + Re \frac{(t_f+t_p)}{t_f} v_f^* \frac{\partial u_f^*}{\partial Y^*} = -Re \frac{(t_f+t_p)^2}{t_f \cdot L} \frac{\partial P_f^*}{\partial X^*} + \frac{(t_f+t_p)^2}{L^2} \frac{\partial^2 u_f^*}{\partial X^{*2}} + \frac{\partial^2 u_f^*}{\partial Y^{*2}} \quad (1)$$

$$Re \frac{(t_f+t_p)^2}{t_f \cdot L} u_f^* \frac{\partial v_f^*}{\partial X^*} + Re \frac{(t_f+t_p)}{t_f} v_f^* \frac{\partial v_f^*}{\partial Y^*} = -Re \frac{(t_f+t_p)}{L} \frac{\partial P_f^*}{\partial Y^*} + \frac{(t_f+t_p)^2}{L^2} \frac{\partial v_f^{*2}}{\partial X^{*2}} + \frac{\partial v_f^{*2}}{\partial Y^{*2}} \quad (2)$$

$$-\frac{kRe}{t_f L} \frac{\partial \langle P_p \rangle^*}{\partial X^*} + \frac{k}{\varepsilon L^2} \frac{\partial^2 \langle u_p \rangle^*}{\partial X^{*2}} + \frac{k}{\varepsilon(t_f+t_p)^2} \frac{\partial^2 \langle u_p \rangle^*}{\partial Y^{*2}} - \langle u_p \rangle^* = 0 \quad (3)$$

$$-\frac{kRe}{t_f(t_f+t_p)} \frac{\partial \langle p_p \rangle^*}{\partial Y^*} + \frac{k}{\varepsilon L^2} \frac{\partial^2 \langle v_p \rangle^*}{\partial X^{*2}} + \frac{k}{\varepsilon(t_f+t_p)^2} \frac{\partial^2 \langle v_p \rangle^*}{\partial Y^{*2}} - \langle v_p \rangle^* = 0 \quad (4)$$

Where, the non-dimensional parameters are defined as: $X^*$ ($X\,L^{-1}$), $Y^*$ ($Y\,(t_f+t_p)^{-1}$), $u_f^*$ ($u_f\,u_{in}^{-1}$), $v_f^*$ ($v_f\,u_{in}^{-1}$), $P_f^*$ ($P_f\,(\rho u_{in}^2)^{-1}$), $\langle u_p \rangle^*$ ($\langle u_p \rangle\,u_{in}^{-1}$), $\langle v_p \rangle^*$ ($\langle v_p \rangle\,u_{in}^{-1}$), $\langle P_p \rangle^*$ ($\langle P_p \rangle\,(\rho u_{in}^2)^{-1}$) and $Re$

$((\rho u_{in} t_f/\mu))$. $t_f$ is the length of flow channel, $t_p$ is the thickness of flow channel, $L$ is the length of flow channel, $k$ is the permeability of porous electrode and $\varepsilon$ is the porosity of porous electrode.

## 3. Simulation Methods and Parameters

The mathematical model is solved in the multi-physics CFD package COMSOL 3.5 a software together with self-written MATLAB programing codes. The non-linear PARDISO solver is adopted. The type of mesh employed is advancing front triangular unstructured grid. The importance of the mesh density around the interface between the flow channel and porous electrode leads to a refinement of the grids. An example of refined mesh for the part of geometry with one layer of carbon electrode is shown in Fig. 2 (c). The grids independent analysis was made to guarantee the mesh quality. The maximum number of refined grids is 416,840. The relative tolerance is $10^{-6}$. All the results presented here are independent with mesh number. Two-dimensional physical parameters for dimensions, properties and initial operation conditions are given in Table 1 [20,22-24]. The thickness, width and length for a single passage of the serpentine flow channel are measured corresponding to be 0.1 cm, 0.1 cm and 2 cm, respectively. The thickness for a single layer of 10 AA SGL carbon fiber paper electrode is ~0.041 cm. The entrance volumetric flow rate $Q_{in}$=20 cm$^3$ min$^{-1}$ gives $u_{in}$=33.3 cm s$^{-1}$ and it is based on the cross section dimension of the flow channel. Compared with classic electrochemical engineering reactors [29-34], there is no turbulence promoter with separation "S" term involved in our VRFB with flow fields. The width (B) of electrode simulated in one segment of flow channel over electrode layer is 0.1 cm. The Reynolds number of electrolyte flow is 91.5 based on the entrance volumetric flow rate of 20 cm$^3$ min$^{-1}$ and properties of electrolyte as shown in Table 1.

The literature on the permeability of typical carbon fiber electrode is limited. Gostick et al. [35] pointed out that this value could be $1\times10^{-11}$ m$^2$ while the one reported by Weber et al. [1] was $2\times10^{-11}$ m$^2$. Nevertheless, the permeability is estimated to be $2.31\times10^{-10}$ m$^2$ through using Carman-Kozney model [36,37]. In the last section, the effect of permeability on maximum current density brought by the volumetric flow penetration through the porous electrode will be discussed. As matter of fact, the degradations [38,39] of carbon paper electrode or carbon felts (typical electrodes used in practical vanadium flow batteries) are hard to be avoidable during the flow battery cycling. The degradation mechanisms of electrodes used in VRFB on their conductivities and permeability are beyond the scope of this study. The permeability of electrode layer is assumed to be invariable in this work. We also have examined the effect of permeability on volumetric flow penetration and maximum current density in later sections 4.3 and 4.4.

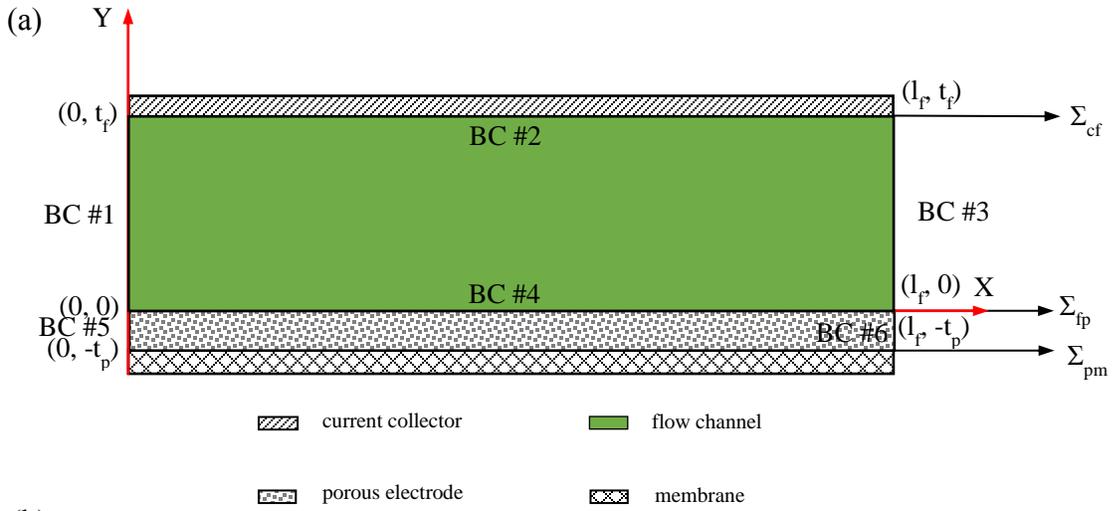

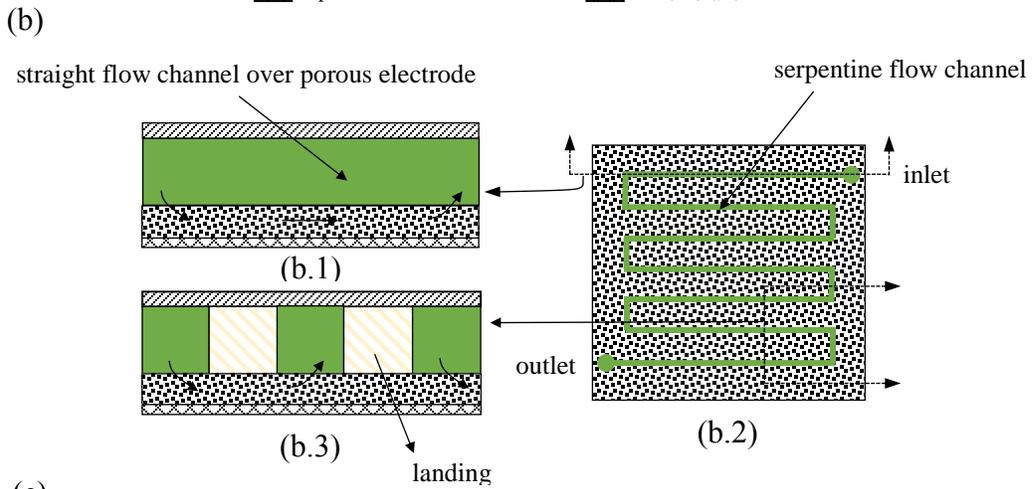

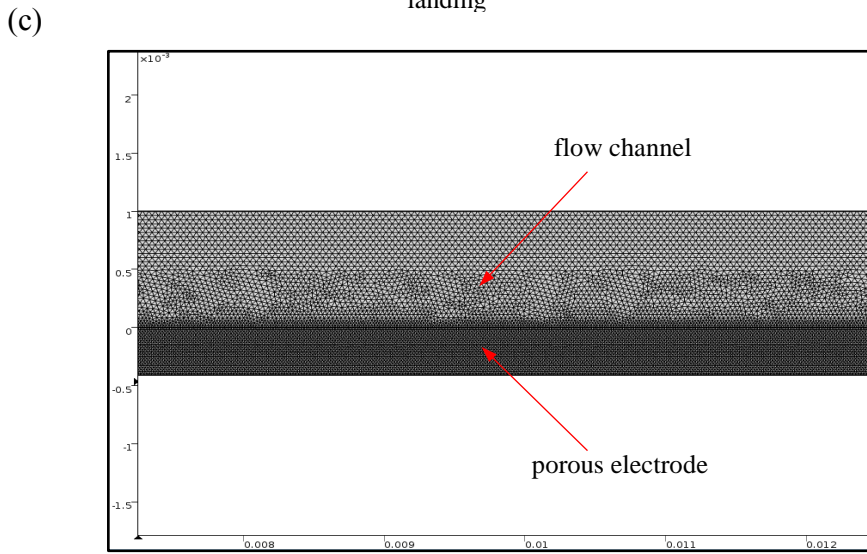

**Fig. 2.** (a) Two-dimensional model of flowing through the straight flow channel and over the porous electrode in the RFBs with flow field (not drawn to scale); (b) three-dimensional model (not drawn to scale): (b.1) straight flow channel over porous electrode, (b.2) serpentine flow channel over porous electrode and (b.3) landing channel view of (b.2) and (c) an example of refined mesh for a part of the flow channel and porous electrode domains of (a)

The continuities of flow velocity and normal stress are applied at the interface boundary between the flow channel and porous electrode (*BC* #4, see Fig. 2 (a)) [40-44]. The flow velocity and pressure type are defined at *BC* #1 (see Fig. 2 (a)) and *BC* #3 (see Fig. 2 (a)), respectively. *BC* #2 (see Fig. 2 (a)), *BC* #5 (see Fig. 2 (a)) and *BC* #6 (see Fig. 2 (a)) are set as "no slip" boundary conditions. Under the realistic flow cell operation condition in the experiments, plastic tubes are used to connect the tank electrolyte reservoirs and holes in the graphite plates engraved with serpentine flow fields. The layouts of holes and serpentine flow fields are relied on the designs and constructions of flow cell hardware. The possible designs for the incoming electrolyte flow direction in the hole is either perpendicular (see Fig. 3 (a) or (d) and (b) or (e)) or parallel (see Fig. 3 (c) or (f)) to the inlet of serpentine flow field. The only difference between Fig. 3 (a) or (d) and (b) or (e) is the head flow channel. For the case of Fig. 3 (a) or (d), this incoming electrolyte flow directly goes into the flow field after 90 degree turn from the perpendicularly oriented feeder tube. The electrolyte flow does not have chance to develop and this type of flow is more likely to be uniform [45]. This uniform flow is treated as ideal plug flow inlet condition for the electrolyte flow through the serpentine flow field over electrode layer. Nevertheless, for the case of Fig. 3 (b) or (e), the electrolyte flow has a chance to develop in a head flow channel prior to flow through the serpentine flow field-electrode layer system and this type of flow is more likely to be parabolic [45]. This flow mode is considered as ideal parabolic flow inlet condition. The case of Fig. 3 (c) or (f) is not a typical one for the experimental flow cell hardware. Under the parallel flow condition, the incoming electrolyte flow has a chance to develop through the plastic tube without any 90 degree turn from the plastic feeder tube. It is likely that the fully developed regime is approached. This type of flow condition is similar to the one of Fig. 3 (b) or (e) as ideal parabolic flow condition. Both ideal plug flow inlet and ideal parabolic flow inlet conditions are discussed in this study. The inlet flow condition is also affected by the rheology of electrolyte. For example, the plug flow inlet condition will occur when electrolyte has a high viscosity (e.g. electrolyte containing particles or suspension or slurry or semi-solid electrolyte) [9,10,46-52]. High viscosity electrolyte can be considered as the non-Newtonian flow and used in the electrochemical flow capacitor (EFC). Also, the shear thinning behavior can be observed in the suspension or slurry or semi-solid non-Newtonian electrolyte flow. This transport phenomenon is related to the relationship between electrolyte viscosity and shear rate. Under the condition of increasing viscosity vs. decreasing shear rate, the shear thinning behavior occurs. Vice versa, the shear thickening behavior occurs under the condition of increasing viscosity with increasing shear rate. The discussions of the non-Newtonian electrolyte flow is beyond the scope of this study.

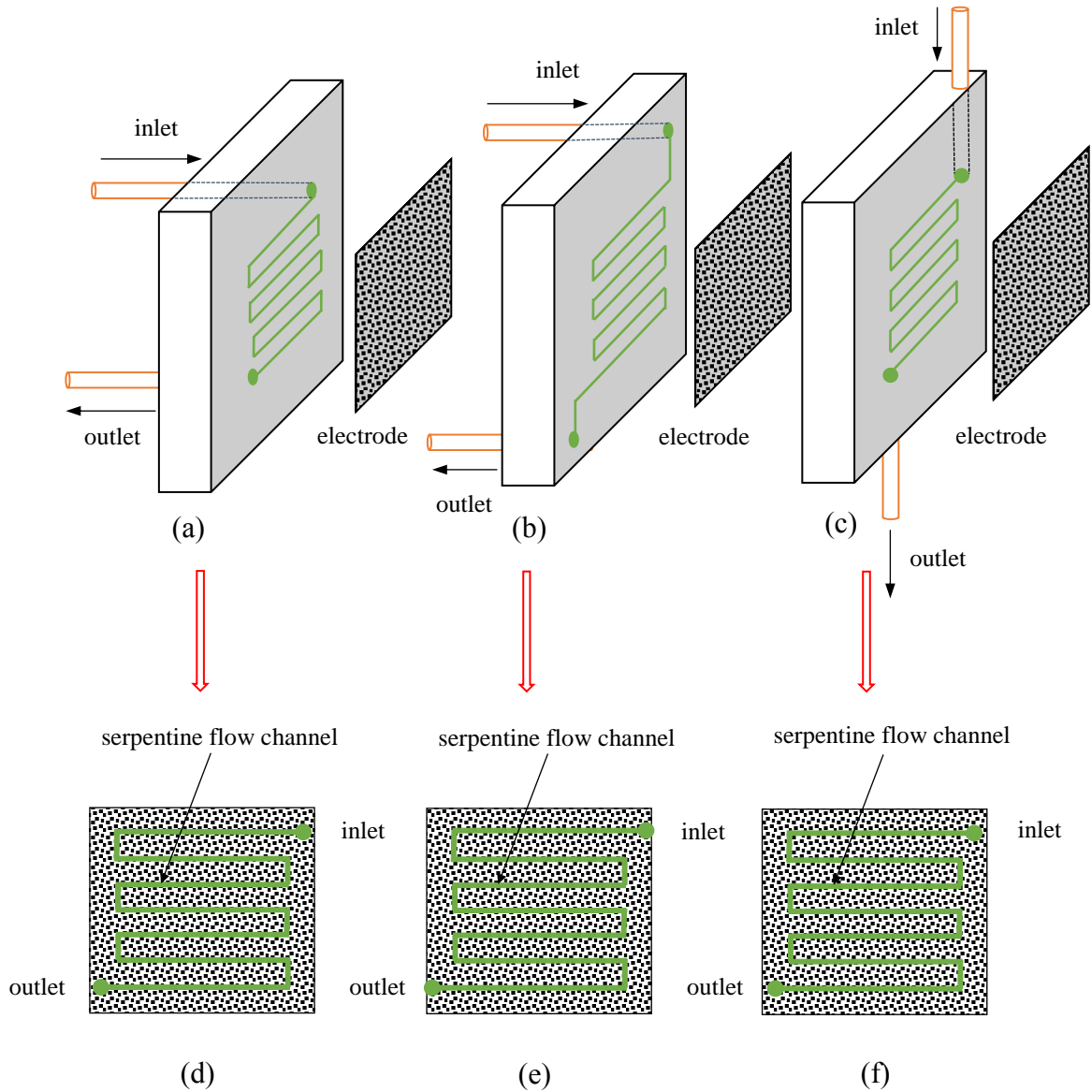

Fig. 3. Schematic diagram of possible electrolyte flow types in RFBs with serpentine flow field: (a) incoming electrolyte flow is perpendicular to serpentine flow field without head flow channel; (b) incoming electrolyte flow is perpendicular to serpentine flow field with head flow channel; (c) incoming electrolyte flow is parallel to serpentine flow field; (d) two-dimensional flow channel-porous electrode view of (a); (e) two-dimensional flow channel-porous electrode view of (b) and (f) two dimensional flow channel-porous electrode view of (c)

# 4. Results and Discussions

## 4.1 Flow patterns

The evolutions of non-dimensional $u_f^*$ ($u_f\, u_{in}^{-1}$) and $<u_p>^*$ ($<u_p>\, u_{in}^{-1}$) given in Table 1 along the normalized longitudinal $Y^*$ ($Y\,(t_f+t_p)^{-1}$) from -0.29 to 0.71 for four different $X^*$ ($X\,L^{-1}$): 0, 0.1, 0.2 and 0.7 as the flow develops from ideal parabolic flow type entrance to developing (the gradient of velocity along $X^*$ ($X\,L^{-1}$) is not zero and in the progress of approaching to be zero) and fully developed (the gradient of velocity along $X^*$ ($X\,L^{-1}$) is zero) regions are shown in Fig. 4. In order to clearly visualize the flow in the porous electrode, the permeability of $9.7\times10^{-9}$ m² is applied. The normalized average velocity $u_{f\,avg}^*$ defined in Eq. (5) [22,23] decreases from 1 to 0.94 while $<u_p>^*_{avg}$ defined in Eq. (6) [22,23] shows an opposite tendency that increases from 0 to 0.15 as the fully developed region is approached.

$$\left(u_f^*{}_{avg}\right)_{X^*} = \frac{t_f+t_p}{t_f}\int_0^{\frac{t_f}{t_f+t_p}} u_f^*(X^*,Y^*)dY^* \tag{5}$$

$$\left(\langle u_p\rangle^*{}_{avg}\right)_{X^*} = \frac{t_f+t_p}{t_p}\int_{-\frac{t_p}{t_f+t_p}}^{0} \langle u_p\rangle^*(X^*,Y^*)dY^* \tag{6}$$

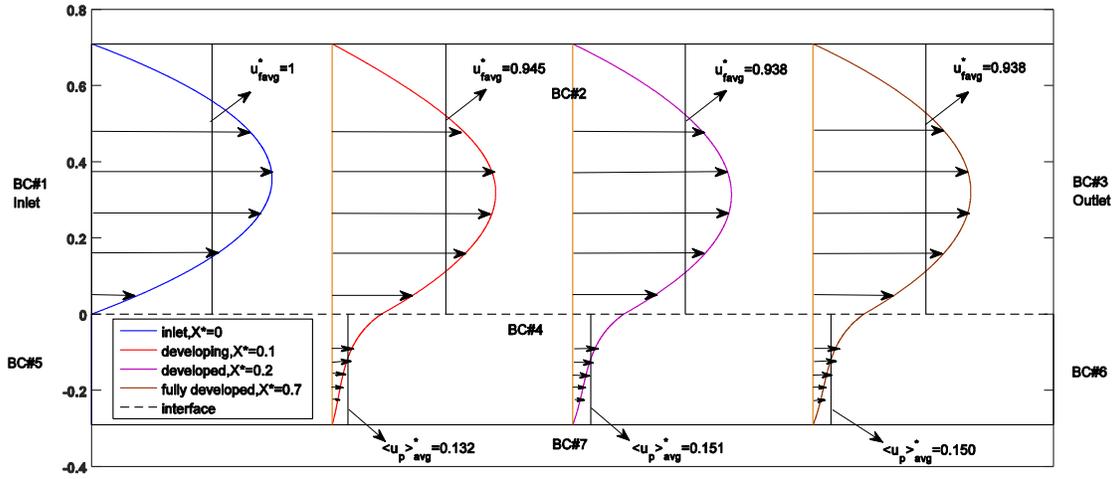

**Fig. 4.** Non-dimensional flow patterns for $u_f^*$ ($u_f\, u_{in}^{-1}$) and $<u_p>^*$ ($<u_p>\, u_{in}^{-1}$) with $Y^*$ ($Y\,(t_f+t_p)^{-1}$) from ideal parabolic flow inlet to developing and fully developed regions in the flow channel and porous electrode at four $X^*$ ($X\,L^{-1}$): 0, 0.1, 0.2 and 0.7 ($k=9.7\times10^{-9}\,m^2$, $\varepsilon=0.78$, $Re=91.5$)

### 4.2 Flow and pressure distributions

The flow distributions under the conditions of ideal plug flow inlet and ideal parabolic flow inlet are studied. In our previous publication (Ref. [23]), a shoot (means a dramatic increase) was found in the flow velocity distributions in the porous layer with a small $X^*$ ($X\,L^{-1}$) when the ideal plug flow inlet condition was applied. A possible reason is that the interface forwarding electrolyte flow momentum through the flow channel force the electrolyte with a fast penetration near to the left wall (*BC* #5, see Fig. 2 (a)) of the porous electrode at the interface under the ideal plug flow inlet condition. Nevertheless, the shoot disappears under the ideal parabolic flow inlet condition. The normalized flow velocity $<u_p>^*$ ($<u_p>\, u_{in}^{-1}$) distributions along $X^*$ ($X\,L^{-1}$) for different $Y^*$ ($Y\,(t_f+t_p)^{-1}$) in the porous electrode is shown in Fig. 5. It is found that almost no shoot occurs in Fig. 5 (b) under the ideal parabolic flow inlet condition. A possible explanation is that the initial entrance interface velocity is zero, and no large flow momentum could drag the flow in the flow channel with a sharp penetration into the porous electrode under the ideal parabolic flow inlet condition.

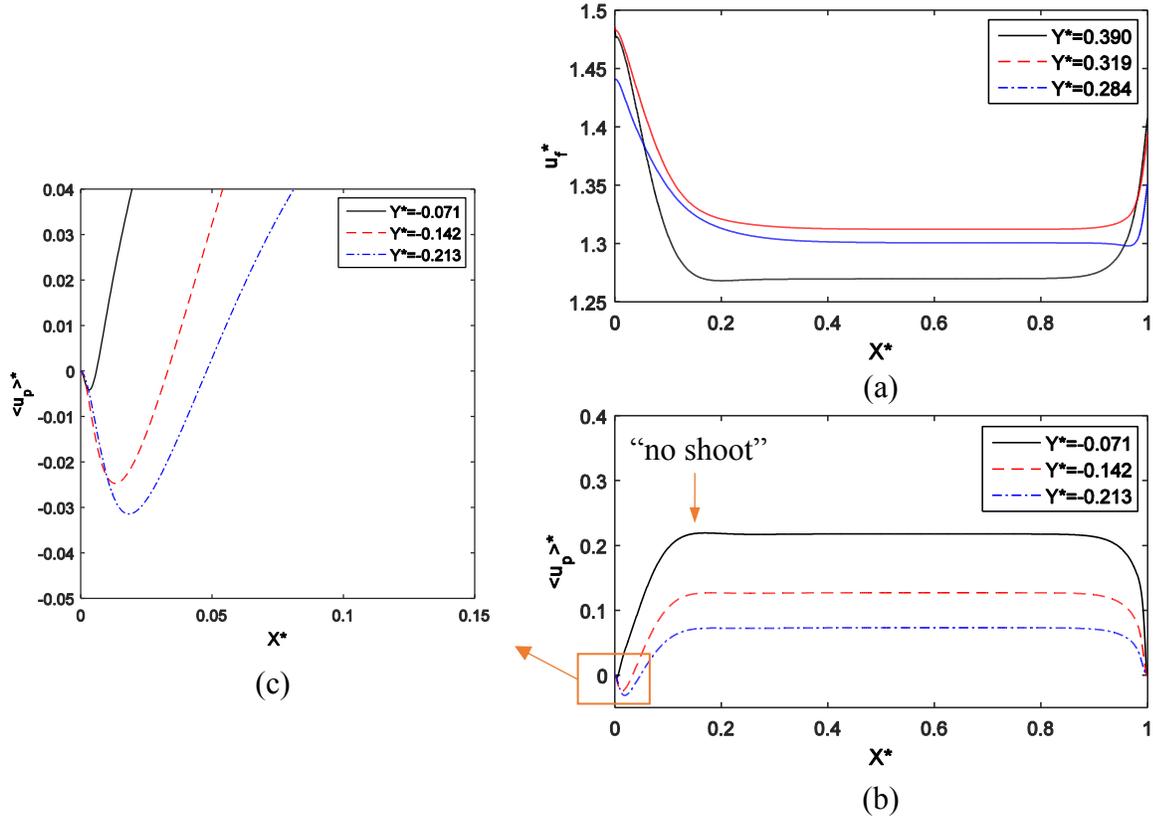

**Fig. 5.** The trend of normalized $u_f^*$ ($u_f\, u_{in}^{-1}$) and $<u_p>^*$ ($<u_p>\, u_{in}^{-1}$) with $X^*$ ($X\, L^{-1}$) from 0 to 1 at six $Y^*$ ($Y\,(t_f+t_p)^{-1}$): 0.390, 0.319, 0.284, -0.071, -0.142 and -0.213 in the flow channel and electrode under ideal parabolic flow inlet condition ("no shoot")

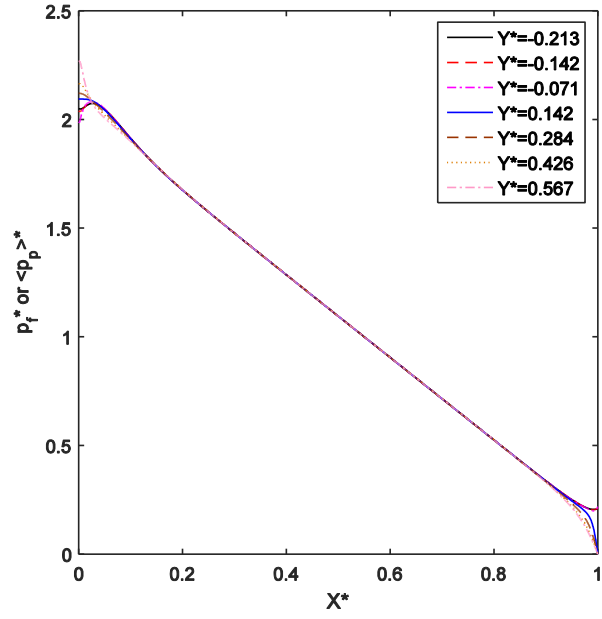

(a)

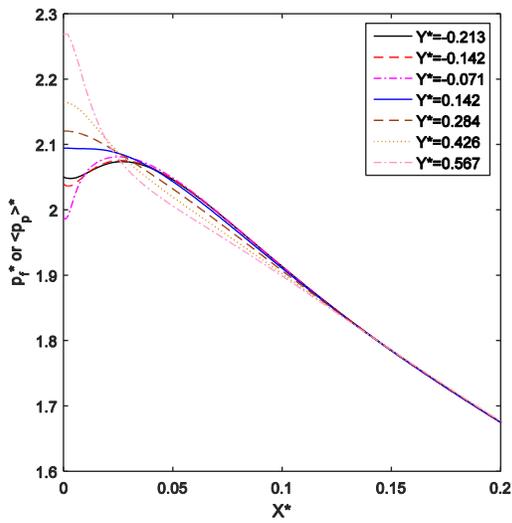

(b)

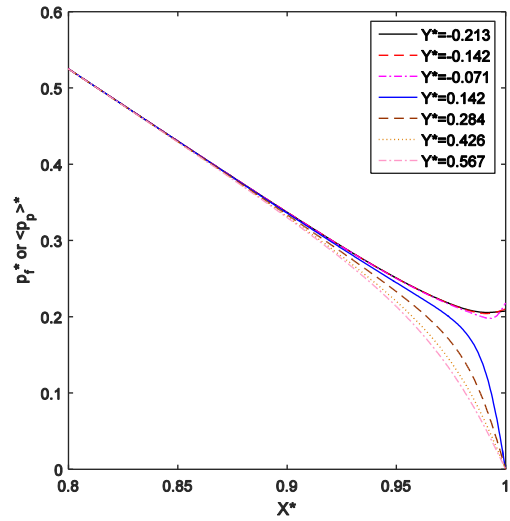

(c)

**Fig.6.** Dimensionless $P_f^*$ ($P_f (\rho u_{in}^2)^{-1}$) and $<P_p>^*$ ($<P_p> (\rho u_{in}^2)^{-1}$) distributions along $X^*$ ($X L^{-1}$) at seven $Y^*$ ($Y (t_f+t_p)^{-1}$): -0.213, -0.142, -0.071, 0.142, 0.284, 0.426 and 0.567; (b) and (c) are the corresponding enlarged local curves of Fig. 4 (a)

The shoot phenomenon also can be explained through the pressure distributions as shown in Fig. 6. The normalized pressure distribution along $X^*$ ($X L^{-1}$) in the flow channel and the porous electrode at seven different $Y^*$ ($Y (t_f+t_p)^{-1}$) as Reynolds number is 91.5 under the ideal plug flow inlet condition is presented. Fig. 6 (b) and (c) are the enlarged local curves of Fig. 6 (a). As $X^*$ ($X L^{-1}$) ranges from ~0.15 to ~0.85, the non-dimensional $P_f^*$ ($P_f (\rho u_{in}^2)^{-1}$) is almost equal to $<P_p>^*$ ($<P_p> (\rho u_{in}^2)^{-1}$), where the fully developed regime is approached. $P_f^*$ ($P_f (\rho u_{in}^2)^{-1}$) is larger than $<P_p>^*$ ($<P_p> (\rho u_{in}^2)^{-1}$) when $X^*$ ($X L^{-1}$) is from 0 to ~0.03 at the beginning of the flow, then electrolyte fluid is driven into the porous electrode from the flow channel. When $X^*$ ($X L^{-1}$) is from ~0.03 to ~0.15, $P_f^*$ ($P_f (\rho u_{in}^2)^{-1}$) is slightly smaller than $<P_p>^*$ ($<P_p> (\rho u_{in}^2)^{-1}$), some flow is forced back into the flow channel from the porous electrode. At the end of the flow, the pressure in the porous layer is larger than in the flow channel as shown in Fig. 6 (c), and it can drive the flow back into the flow channel from the porous electrode. Nevertheless, no electrolyte flow is driven out at the beginning of the flow under the ideal parabolic flow inlet condition. Fig. 5 also reveals that some back flow occurs and it can be explained through the non-dimensional pressure distributions as shown in Fig. 5.

### 4.3 Interface flow and volumetric flow penetration

The distributions of flow velocity $v$ along the $Y$ direction at the interface between the flow channel and porous electrode under the ideal plug flow and the ideal parabolic flow inlet conditions are shown in Fig. 7 (a) and (b), respectively. At most part of the interface between the flow channel and porous electrode, except at the beginning and end of the flow, $v$ velocity is almost zero. The reason is that the flow entrance length under the ideal plug flow inlet condition is ~ 0.4 cm to 0.5 cm as shown in Fig. 8 (b) based on the initial mean velocity of 33.3 cm s$^{-1}$ or the initial entrance volumetric flow rate of 20 cm$^3$ min$^{-1}$. The entrance length is even shorter under the ideal parabolic flow inlet condition as shown in Fig. 8 (d). The total length of the flow channel is 2 cm and this means that at least around 75% to 80% region of the flow channel and porous electrode are in the fully developed regime. Under this case, there is no pressure difference between the flow channel and porous electrode as the electrolyte flow comes to the fully developed regime. Flow penetration cannot occur at the interface without pressure driven force.

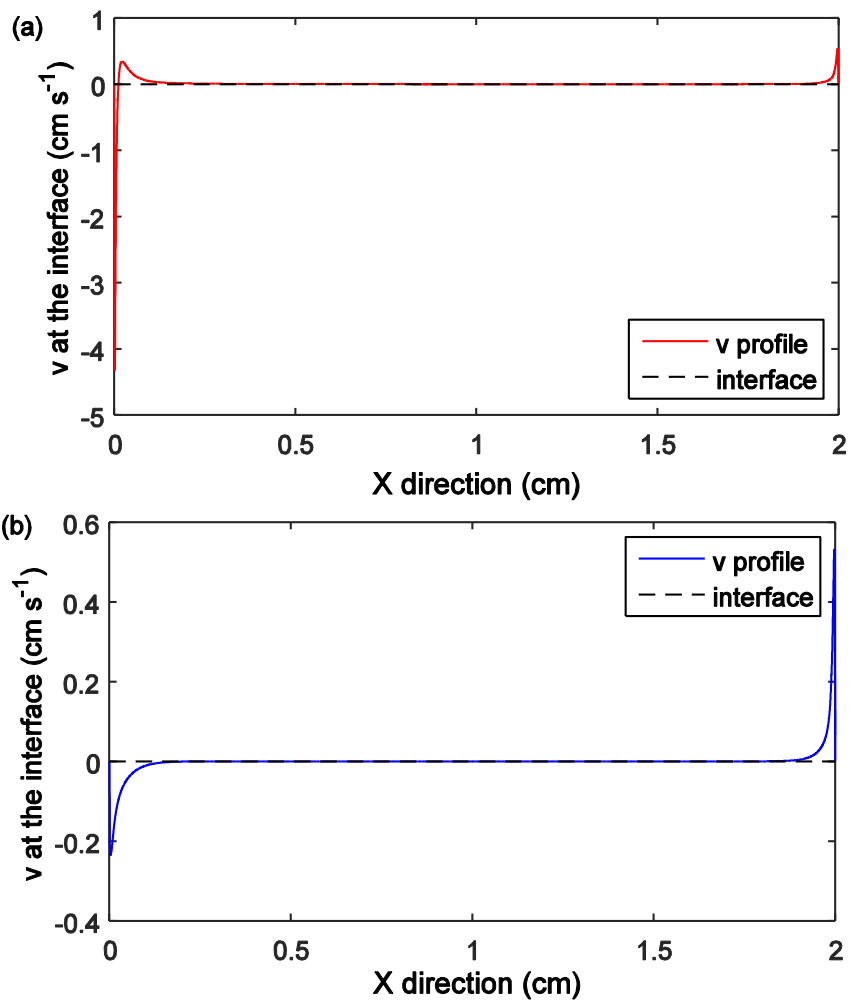

**Fig. 7.** *v* velocity distributions at the interface between the flow channel and porous electrode: (a) ideal plug flow inlet condition and (b) ideal parabolic flow inlet condition (single layer of 10 AA carbon fiber paper, $Q_{in}$=20 cm$^3$ min$^{-1}$)

The enlarged local curves for the beginning and end of the flow as shown in Fig. 7 can be visualized through Fig. 8 (a)-(e). It is found that electrolyte flow penetrates into the porous electrode, then is pushed back into the flow channel at the beginning of the flow under the ideal plug flow inlet condition as shown in Fig. 8 (a)-(c) while the electrolyte flow just penetrates into the porous layer at the beginning of the flow as shown in Fig. 8 (d) and (e) under the ideal parabolic flow inlet condition. At the end of flow, all the flow is driven back into the flow channel and this is because of a no slip wall boundary condition (*BC* #6, see Fig. 2 (a)).

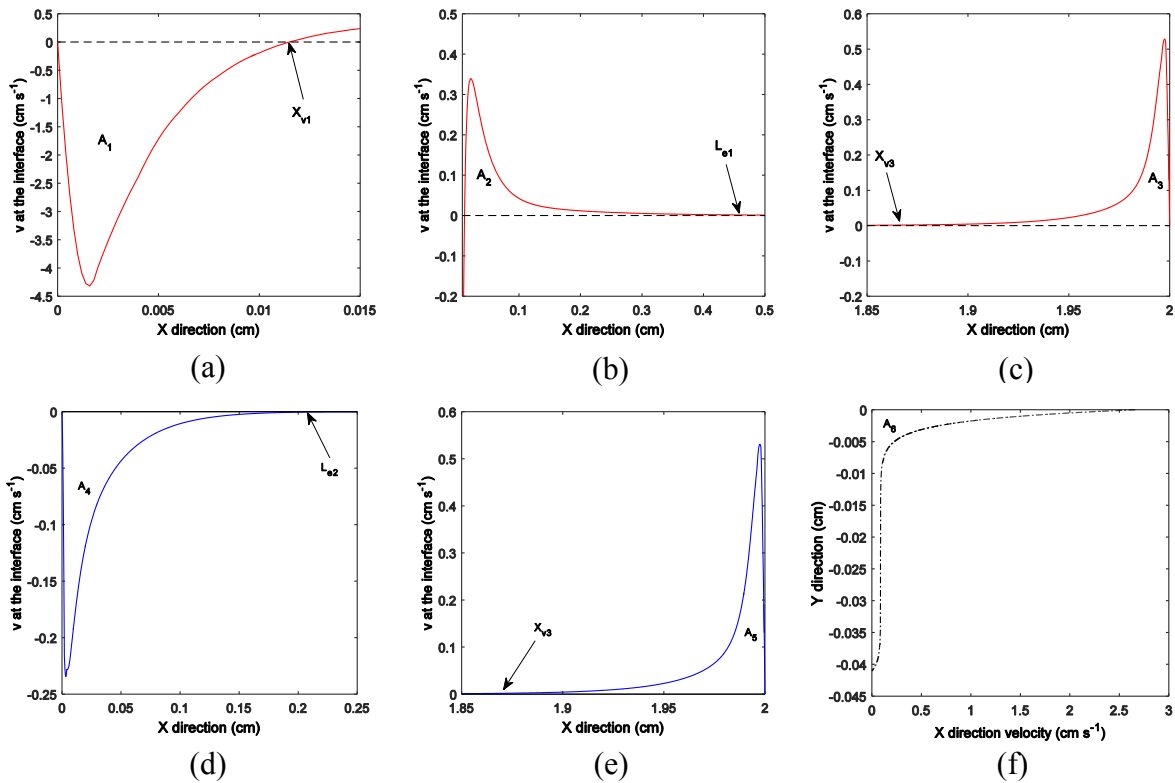

**Fig. 8.** Enlarged local curves of Fig. 7 for the *v* velocity distributions at the beginning and the end of flow: (a), (b) and (c) under ideal plug flow inlet condition; (d) and (e) under ideal parabolic flow inlet condition and (f) *u* velocity distributions as the fully developed regime is approached (single layer of 10 AA carbon fiber paper, $Q_{in}$=20 ml min$^{-1}$)

Although the flow penetration can be output directly from the model, more details on how is the flow penetration through the interface between the flow channel and porous electrode calculated under both the ideal plug flow inlet and ideal parabolic flow inlet boundary conditions are explained. The flow penetration through the interface is the integral of velocity that is perpendicular to the interface (*v* velocity). Under the ideal plug flow inlet condition, the electrolyte flow sharply penetrates into the electrode at the very beginning and then penetrates out with a little bit amount. At the end of flow, all electrolyte is penetrated out. It can be seen that there are three regions ($A_1$, $A_2$ and $A_3$) of local flow penetration (with a "shoot") for the electrolyte flow under the ideal plug flow inlet condition. However, only two regions ($A_4$ and $A_5$) of local flow penetration (without a "shoot") occur under the ideal parabolic flow inlet condition. Under the fully developed regime, no flow penetration occurs. Five piecewise functions of $A_1$, $A_2$, $A_3$, $A_4$ and $A_5$ as shown in Fig. 8 (a)-(e) are defined

$$A_1 = \int_0^{X_{v1}} (v_p)_{\Sigma_{fp}} dX \tag{7}$$

$$A_2 = \int_{X_{v1}}^{L_{e1}} (v_p)_{\Sigma_{fp}} dX \tag{8}$$

$$A_3 = \int_{X_{v2}}^{L} (v_p)_{\Sigma_{fp}} dX \tag{9}$$

$$A_4 = \int_{0}^{L_{e2}} (v_p)_{\Sigma_{fp}} dX \tag{10}$$

$$A_5 = \int_{X_{v3}}^{L} (v_p)_{\Sigma_{fp}} dX \tag{11}$$

Where, $X_{v1}$ is the length from the entrance flow to the flow starting to penetrate out from the porous electrode into the flow channel at the interface at the beginning of flow under the ideal plug flow inlet condition; $L_{e1}$ is the entrance length under the ideal plug flow inlet condition; $X_{v2}$ is the length from the entrance flow to the flow second time penetrating out from the porous electrode at the interface under the ideal plug flow inlet condition; $L_{e2}$ is the entrance length under the ideal parabolic flow inlet condition and $X_{v3}$ is the length from the entrance flow to the flow starting to penetrate out from the porous electrode into the flow channel at the interface under the ideal parabolic flow inlet condition. $A_1$, $A_2$, $A_3$, $A_4$ and $A_5$ are the integrations of $v$ velocity along the interface between the flow channel and porous electrode. The conservation of volumetric flow rate gives the value of ($A_2$-$A_1$) and it is equal to $A_3$, -$A_4$ and $A_5$. Moreover, the volumetric flow rate within the porous electrode is calculated by the integration through the interface is also equal to the integration of $X$ direction velocity $<u_p>$ along the $Y$ direction in the porous electrode (see $A_6$) when fully developed flow regime is approached as shown in Fig. 8 (f). The formula of $A_6$ is given [22-24]

$$A_6 = \int_{-t_p}^{0} (\langle u_p \rangle)_{fd} dY \tag{12}$$

Therefore, the volumetric flow penetration through the porous electrode $(Q_p)_{fc}$ under the flow channel advection as shown in Eq. (13) is equal to $w_{fc}(A_2-A_1)$, $w_{fc}A_3$, $-w_{fc}A_4$, $w_{fc}A_5$ or $w_{fc}A_6$

$$(Q_p)_{fc} = \begin{cases} w_{fc}(A_2 - A_1) \\ w_{fc}A_3 \\ -w_{fc}A_4 \\ w_{fc}A_5 \\ w_{fc}A_6 \end{cases} \tag{13}$$

Here, $(Q_p)_{fc}$ is the volumetric flow penetration through the porous electrode beneath the flow channel. As the electrolyte flow approaches from the initial inlet type to fully developed regime, the implicit form of $(Q_p)_{fc}$ is discussed through Eq. (7) to Eq. (13). More importantly, the landing/rib advection can force much more volumetric flow penetration through the porous electrode. For the segment of one flow channel and one landing/rib, the flow penetration under the rib advection can be estimated (see appendix section)

$$(Q_p)_{lc} \sim \frac{L}{r}(Q_p)_{fc} \tag{14}$$

Where, $r$ is the distance between the adjacent flow channels or length of landing/rib and $L$ is the length of flow channel. In the experimental measurement, $r$ and $L$ is corresponding to be 0.1 cm and 2 cm. It is clear from Eq. (14) that the order magnitude of volumetric flow penetration through the porous electrode under the landing/rib advection is ~ 20 times to be the one under the flow channel advection. The total volumetric flow penetration through the porous electrode under both advections by flow channel and landing/rib is estimated (see appendix section)

$$(Q_p)_{tot} = (Q_p)_{fc} + (Q_p)_{lc} \sim (1 + \frac{L}{r})(Q_p)_{fc} \tag{15}$$

The total volumetric flow penetration through the porous electrode can be related to the dimensions of the flow cell (e.g. length, width and thickness of flow channel and porous electrode), properties of electrolyte (e.g. density and dynamic viscosity), properties of porous electrode (e.g. porosity and permeability), initial ion concentration and entrance volumetric flow rate (or entrance mean linear velocity).

**4.4 Discussions on estimated maximum current density**

The second law of Faraday's electrolysis and volumetric flow balance through the porous electrode yield that the maximum current density can be achieved in RFBs with flow field design limited by the 100% amount of electrolyte reactant consumption through the porous electrode under both flow channel and landing/rib advections yields

$$i_{max} = \frac{nFc(Q_p)_{tot}}{(w_{fc}+r)L} \tag{16}$$

Where, $n$ is the number of electron transferred in the chemical reactions ($n$=1, vanadium flow battery), $F$ is the Faraday constant (96,485 C mol$^{-1}$), $c$ is the bulk concentration (0.001 mol cm$^{-3}$), $(Q_p)_{tot}$ is the total volumetric flow rate through the porous electrode as shown in Eq. (14), $w_{fc}$ is the width of the flow channel, $r$ is the distance between adjacent flow channels or length of landing/rib and $L$ is the length of flow channel. Based on the permeability of 2.31×10$^{-6}$ cm$^2$ estimated by Kozeny-Carman model [36,37], the corresponding theoretical maximum current density estimated by Eq. (16) is 7,917 mA cm$^{-2}$ (vs. experimental result of ~400 mA cm$^{-2}$, [20]) and 15,204 mA cm$^{-2}$ (vs. experimental result of ~750 mA cm$^{-2}$, [14]) for one layer and three layers of SGL 10AA carbon paper electrode. However, Weber et al. [1] ($k=2\times10^{-7}$ cm$^2$) and Gostick et al. [35] ($k=1\times10^{-7}$ cm$^2$) pointed out that the permeability for the typical carbon paper electrode could be the order of $10^{-11}$ m$^2$. It is likely that the permeability estimated by Kozeny-Carman is over-estimated. If the permeability of $1\times10^{-7}$ cm$^2$ is applied, then the corresponding maximum current density estimated by Eq. (16) is 343 mA cm$^{-2}$ and 658 mA cm$^{-2}$ for one layer and three layers of SGL 10AA carbon paper electrode. It is clear that the theoretical maximum current density achieved is sensitive to the value of permeability. The discrepancy between the theoretical maximum current density and experimental value can be caused by the accuracy of parameters (such as permeability, dynamic viscosity, etc.) as shown in Table 1. The full serpentine flow dynamic pattern in the three-dimensional model of serpentine flow channel over the porous electrode will be explored in the future. The further work on mathematical optimizations of flow batteries with different flow field designs is still needed. In this work, we explore flow penetration and its related maximum current density in vanadium flow battery with flow field design over carbon paper electrode. Compared with classic one without flow field design through the carbon felts, two possible benefits brought by using flow field design within improving electrochemical performance: (1) lower ohmic loss in carbon paper electrode (e.g. hundred microns) compared with carbon felts (e.g. several millimeters) and (2) stronger mass transfer through the electrode layer from the flow field caused by the forced convection. Nevertheless, one significant drawback for the flow cell with flow field manifolds is the issue of non-uniform flow reactant distribution within the electrode. A "dead zone" for electrolyte flow always occurs at the flow cell with a bad flow field design. Thus, the optimizations of flow distributions for flow cells with flow field designs by mathematical modeling approaches are desired. The understanding of transport mechanism in the flow cell can be achieved through both experiments and mathematical modeling. Typically, multiphysics modeling analysis is always much cheaper and quicker compared with experimental studies.

## Concluding remarks

A simplified two-dimensional mathematical model is developed to study the electrolyte flow patterns and volumetric flow penetrations in the flow channel over the carbon fiber paper electrode layered system in vanadium flow battery with serpentine flow field design. It is found that the interface flow distributions is different under the ideal plug flow inlet and ideal parabolic flow inlet conditions and this is caused by non-linear pressure behaviors in the flow channel and porous electrode at the beginning and end of flow. The mathematical model of volumetric flow penetration through the integration of interface flow velocity reveals that the volumetric flow penetration within the porous electrode by the flow channel advection is equal under both ideal plug flow and ideal parabolic flow inlet conditions. Flow penetrations under both effects of flow channel and landing/rib advections are estimated. The maximum current density achieved in the flow battery can be predicted based on the 100% amount of electrolyte flow reactant consumption through the porous electrode beneath both flow channel and landing/rib. It is found that the corresponding maximum current density estimated is sensitive to the value of permeability. Based on a proper permeability, the theoretical maximum current density have reasonable agreement with experimental results of vanadium flow battery with one and three layers SGL 10AA carbon paper electrode.

**Acknowledgements**

This work is partially supported by the flow battery project (DE-AR0000352) funded from Department of Energy (DOE) of the United States. We thank Dr. D.L. Feke and Dr. P. Barnhart from Case Western Reserve University for discussing the boundary conditions of ideal plug flow and ideal parabolic flow inlet boundary conditions. We thank Dr. N.C. Hoyt from Case Western Reserve University for give the suggestions on the volumetric flow rate through the porous electrode. We gratefully acknowledge the helpful advice from Dr. A.Z. Weber from Lawrence Berkeley National Laboratory and Dr. M.M. Mench from University of Tennessee at Knoxville.

**Appendix**

The effect of landing/rib on the flow penetration in the actual three-dimensional model is incorporated in this two-dimensional model through the scaling analysis. The derivation process of Eqs. (14) and (15) are explained in this appendix section. The serpentine flow field over electrode layer can be considered as several periodic segments of two adjacent flow channels (#1 and #2) and one landing/rib channel over the electrode layer. This is illustrated in Fig. A1: (a) #1 flow channel over the electrode; (b) one landing channel over the electrode and (c) #2 flow channel over the electrode. The mathematical derivation process of average velocity and volumetric flow rate through the electrode is described. For Fig. A1 (a), the average flow velocity and volumetric flow rate through the porous electrode based on Darcy's law yield

$$\langle u_p \rangle_{fc,1} = -k\nabla(P_p)/\mu \sim k\left((P_p)_1 - (P_p)_2\right)/\mu L_p \tag{A.1}$$

$$(Q_p)_{fc,1} = \langle u_p \rangle_{fc,1} w_p t_p \sim k w_p t_p \left((P_p)_1 - (P_p)_2\right)/\mu L_p \tag{A.2}$$

Where, $k$ is the permeability of porous electrode, $\mu$ is the electrolyte dynamic viscosity, $w_p$ is the width of porous electrode, $t_p$ is the thickness of porous electrode and $L_p$ is the length of porous electrode.

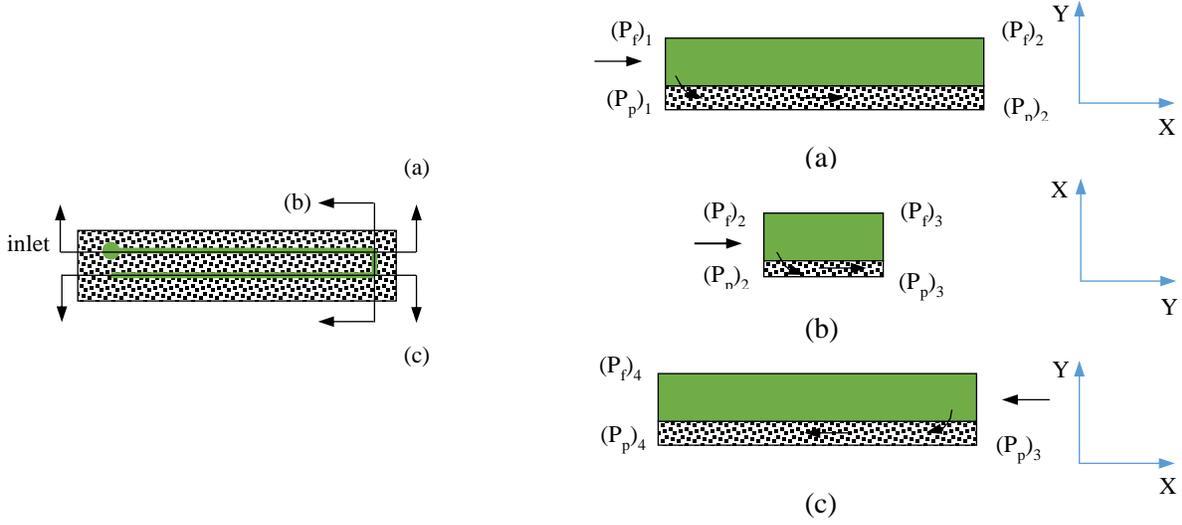

Fig. A1 Schematic diagram of flow penetration through the porous electrode under the effect of landing/rib: (a) one flow channel (#1) over the electrode; (b) one landing channel over the electrode and (c) adjacent flow channel (#2) over the electrode.

For Fig. A1 (b), the average flow velocity and volumetric flow rate through the electrode beneath the landing channel

$$\langle v_p \rangle_{lc} = -k\nabla(P_p)/\mu \sim k\left((P_p)_2 - (P_p)_3\right)/\mu r \tag{A.3}$$

$$(Q_p)_{lc} = \langle v_p \rangle_{lc} L_p t_p \sim k L_p t_p \left((P_p)_2 - (P_p)_3\right)/\mu r \tag{A.4}$$

Where, $r$ is the distance between adjacent flow channels (#1 and #2). For Fig. A1 (c), the average flow velocity and volumetric flow rate through the porous electrode

$$\langle u_p \rangle_{fc,2} = -k\nabla(P_p)/\mu \sim k\left((P_p)_3 - (P_p)_4\right)/\mu L_p \tag{A.5}$$

$$(Q_p)_{fc,2} = \langle u_p \rangle_{fc,2} w_p t_p \sim k w_p t_p \left((P_p)_3 - (P_p)_4\right)/\mu L_p \tag{A.6}$$

In this simulation with a small Reynolds number of 91.5, laminar flow regime is approached as the electrolyte flow through the flow channel and porous electrode. The pressure in the laminar flow always exhibits a linear behavior and this also can be demonstrated by Fig. (6): the pressure in most part region of flow channel and porous electrode is linear. The pressure gradient in (a), (b) and (c) as shown in Fig. A1 can be approximated to be equal

$$\left((P_p)_1 - (P_p)_2\right)/L_p \sim \left((P_p)_2 - (P_p)_3\right)/r \sim \left((P_p)_3 - (P_p)_4\right)/L_p \tag{A.7}$$

Thus, Eqs. (A.2), (A.4), (A.6), (A.7) yield

$$(Q_p)_{fc,1} \sim (Q_p)_{fc,2} \sim (Q_p)_{lc} r/L_p \tag{A.8}$$

# References

[1] A.Z. Weber, M.M. Mench, J.P. Meyers, P.N. Ross, J.T. Gostick, Q. Liu, "Redox flow batteries: a review", J. Appl. Electrochem. 41 (2011) 1137-1164.

[2] W. Wang, Q. Luo, B. Li, X. Wei, L. Li, Z. Yang, "Recent progress in redox flow battery research and development", Adv. Funct. Mater. 23 (2013) 970-986.

[3] A. Parasuraman, T.M. Lim, C. Menictas, M. Skyllas-Kazacos, "Review of material research and development for vanadium redox flow battery applications", Electrochim. Acta 101 (2013) 27-40.

[4] P. Alotto, M. Guarnieri, F. Moro, "Redox flow batteries for the storage of renewable energy: a review", Renew. Sust. Energy Rev. 29 (2014) 325-335.

[5] M. Skyllas-Kazacos, M. Rychcik, R.G. Robins, A.G. Fane, M.A. Green, "New all-vanadium redox flow cell", J. Electrochem. Soc. 133 (1986) 1057-1058.

[6] Z. Li, W. Dai, L. Yu, J. Xi, X. Qiu, L. Chen, "Sulfonated poly(ether ether ketone)/mesoporous silica hybrid membrane for high performance vanadium redox flow battery", J. Power Sources 257 (2014) 221-229.

[7] L.H. Thaller, "Redox flow cell energy storage systems", Department of Energy, Washington, DC, DOE/NASA/1002-79/3; National Aeronautics and Space Administration: Washington, DC, NASA TM-79143 (1979).

[8] L.W. Hruska, R.F. Savinell, "Investigation of factors affecting performance of the iron-redox battery", J. Electrochem. Soc. 128 (1981) 18-25.

[9] T.J. Petek, N.C. Hoyt, R.F. Savinell, J.S. Wainright, "Characterizing slurry electrodes using electrochemical impedance spectroscopy", J. Electrochem. Soc. 163 (2016) A5001-A5009.

[10] M. Duduta, B. Ho, V.C. Wood, P. Limthongkul, V.E. Brunini, W.C. Carter, Y.M. Chiang, "Semi-solid lithium rechargeable flow battery", Adv. Energy Mater. 1 (2011) 511-516.

[11] L. Li, S. Kim, W. Wang, M. Vijayakumar, Z. Nie, B. Chen, J. Zhang, G. Xia, J. Hu, G. Graff, J. Liu, Z. Yang, "A stable vanadium redox-flow battery with high energy density for large-scale energy storage", Adv. Funct. Mater. 1 (2011) 394-400.

[12] A.A. Shinkle, A.E.S. Sleightholme, L.D. Griffith, L.T. Thompson, C.W. Monroe, "Degradation mechanisms in the non-aqueous vanadium acetylacetonate redox flow battery", J. Power Sources 206 (2012) 490-496.

[13] C. Flox, M. Skoumal, J. Rubio-Garcia, T. Andreu, J.R. Morante, "Strategies for enhancing electrochemical activity of carbon-based electrodes for all-vanadium redox flow batteries", Appl. Energy 109 (2013) 344-351.

[14] A.A. Shah, R. Tangirala, R. Singh, R.G.A. Wills, F.C. Walsh, "A dynamic unit cell model for the all-vanadium flow battery", J. Electrochem. Soc. 158 (2011) A671-A677.

[15] K.W. Knehr, E. Agar, C.R. Dennison, A.R. Kalidindi, E.C. Kumbur, "A transient vanadium flow battery model incorporating vanadium crossover and water transport through the membrane", J. Electrochem. Soc. 159 (2012) A1446-A1459.

[16] J. Newman, C.W. Tobias, "Theoretical analysis of current distribution in porous electrodes", J. Electrochem. Soc. 109 (1962): 1183-1191.

[17] J. Newman, W. Tiedemann, "Porous-electrode theory with battery applications", AlChE Journal 21 (1975): 25-41.

[18] A.A. Shah, M.J. Watt-Smith, F.C. Walsh, "A dynamic performance model for redox-flow batteries involving soluble species", Electrochim. Acta 53 (2008) 8087-8100.

[19] D. You, H. Zhang, J. Chen, "A simple model for the vanadium flow battery", Electrochim. Acta 54 (2009) 6827-6836.

[20] D.S. Aaron, Q. Liu, Z. Tang, G.M. Grim, A.B. Papandrew, A. Turhan, T.A. Zawodzinski, M.M. Mench, "Dramatic performance gains in vanadium redox flow batteries through modified cell architecture", J. Power Sources 206 (2012) 450-453.

[21] D. Aaron Z. Tang, A.B. Papandrew, T.A. Zawodzinski, "Polarization curve analysis of all-vanadium redox flow batteries", J. Appl. Electrochem. 41 (2011) 1175-1182.

[22] X. Ke, "CFD studies on mass transport in redox flow batteries", Master Thesis, Case Western Reserve University, United States, 2014 (https://etd.ohiolink.edu).

[23] X. Ke, J.I.D. Alexander, J.M. Prahl, R.F. Savinell, "Flow distribution and maximum current density studies in redox flow batteries with a single passage of the serpentine flow channel", J. Power Sources 270 (2014) 646-657.

[24] X. Ke, J.I.D. Alexander, J.M. Prahl, R.F. Savinell, "A simple analytical model of coupled single flow channel over porous electrode in vanadium redox flow battery with serpentine flow channel", J. Power Sources 288 (2015) 308-313.

[25] T.J. Latha, S. Jayanti, "Hydrodynamic analysis of flow fields for redox flow battery applications", J. Appl. Electrochem. 44 (2014) 995-1006.

[26] S. Kumar, S. Jayanti, "Effect of flow field on the performance of an all-vanadium redox flow battery", J. Power Sources 307 (2016) 782-787.

[27] W. Merzkirch, "Flow Visualization", Book, 2nd Edition, Academic Press, Inc., United States, 1987.

[28] P.V. Danckwerts, "Continuous flow systems: Distribution of residence times", Chem. Eng. Sci. 2 (1953) 1-13.

[29] C. Ponce-de-León, G.W. Reade, I. Whyte, S.E. Male, F.C. Walsh, "Characterization of the reaction environment in a filter-press redox flow reactor", Electrochim. Acta 52 (2007) 5815-5823.

[30] C. Ponce-de León, I. Whyte, G.W. Reade, S.E. Male, F.C. Walsh, "Mass transport and flow dispersion in the compartments of a modular 10 cell filter-press stack", Aust. J. Chem. 61 (2008) 797-804.

[31] P. Leung, X. Li, C. Ponce de-Léon, L. Berlouis, C.T.J. Low, F.C. Walsh, "Progress in redox flow batteries, remaining challenges and their applications in energy storage", RSC Adv. 2 (2012) 10125 - 10156.

[32] C. Ponce-de-León, A. Frías-Ferrer, J. González-Garcia, D.A. Szánto, F.C. Walsh, "Redox flow cells for energy conversion", J. Power Sources 160 (2006) 716-732.

[33] G. Kear, A.A. Shah, F.C. Walsh, "Development of the all-vanadium redox flow battery for energy storage: a review of technological, financial and policy aspects", Inter. J. Energ. Res. 36 (2012) 1105-1120.

[34] M.J. Watt-Smith, P. Ridley, R.G.A. Wills, A.A. Shah, F.C. Walsh, "The importance of key operational variables and electrolyte monitoring to the performance of an all vanadium redox flow battery", J. Chem. Tech. Biotech. 88 (2013) 126-138.

[35] J.T. Gostick, M.W. Fowler, M.D. Pritzker, M.A. Ioannidis, L.M. Behra, "In-plane and through-plane gas permeability of carbon fiber electrode backing layers", J. Power Sources 162 (2006) 228-238.

[36] P.C. Carman, "Fluid flow-through granular beds", Trans. Inst. Chem. Eng. Lond. 15 (1937) 150-166.

[37] J. Kozeny, "Ueber kapillare leitung des Wassers im Boden", Stizungsber Akad. Wiss. Wien. 136 (1927) 271-306.

[38] P. Trogadas, O.O. Taiwo, B. Tjaden, T.P. Neville, S. Yun, J. Parrondo, V. Raman, M.O. Coppens, D.J.L. Brett, P.R. Shearing, "X-ray micro-tomography as a diagnostic tool for the electrode degradation in vanadium redox flow batteries", Electrochem. Commun. 48 (2014) 155-159.

[39] I. Derr, M. Bruns, J. Langer, A. Fetan, J. Melke, C. Roth, "Degradation of all-vanadium redox flow batteries (VRFB) investigated by electrochemical impedance and X-ray photoelectron spectroscopy: Part 2 electrochemical degradation", J. Power Sources 325 (2016) 351-359.

[40] K. Vafai, R. Thiyagaraja, "Analysis of flow and heat transfer at the interface region of a porous medium", Int. J. Heat and Mass Transfer 30 (1987) 1391-1405.

[41] S.J. Kim, C.Y. Choi, "Convection heat transfer in porous and overlaying layers heated from below", Int. J. Heat and Mass Transfer 39 (1996) 319-329.

[42] D. Poulikakos, M. Kazmierczak, "Forced convection in a duct partially filled with a porous material", J. Heat Transfer 109 (1987) 653-662.

[43] M. Ehrhardt, "An introduction to fluid-porous interface coupling, http://www.math.uni-wuppertal.de", last accessed: May 2014.

[44] V. Laptev, "Numerical solution of coupled flow in plain and porous media", Ph.D. Thesis, University of Kaiserslautern, 2003.

[45] J.A. Schetz, A.E. Fuhs, "Fundamentals of fluid mechanics", Book, 3$^{rd}$ Edition, John Wiley & Sons, United States, 1999.

[46] K.C. Smith, Y.M. Chiang, W.C Carter, "Maximizing energetic efficiency in flow batteries utilizing non-Newtonian Fluids", J. Electrochem. Soc. 161 (2014) A486-A496.

[47] V.E. Brunini, Y.M. Chiang, W.C. Carter, "Modeling the hydrodynamic and electrochemical efficiency of semi-solid flow batteries", Electrochim. Acta 69 (2012) 301-307.

[48] K.C. Smith, V.E. Brunini, Y. Dong, Y.M. Chiang, W.C. Carter, "Electroactive-zone extension in flow-battery stacks", Electrochim. Acta 147 (2012) 460-469.

[49] J.W. Campos, M. Beidaghi, K.B. Hatzell. C.R. Dennison, B. Musci, V. Presser, E.C. Kumbur, Y. Gogotsi, "Investigation of carbon materials for use as a flowable electrode in electrochemical flow capacitors", Electrochim. Acta 98 (2013) 123-130.

[50] K.B. Hatzell, M. Beidaghi, J.W. Campos, C.R. Dennison, E.C. Kumbur, Y. Gogotsi, "A high performance pseudocapacitive suspension electrode for the electrochemical flow capacitor", Electrochim. Acta 111 (2013) 888-897.

[51] C.R. Dennison, M. Beidaghi, K.B. Hatzell, J.W. Campos, Y. Gogotsi, E.C. Kumbur, "Effects of flow cell design on charge percolation and storage in the carbon slurry electrodes of electrochemical flow capacitors", J. Power Sources 247 (2014) 489-496.

[52] M. Boota, K.B. Hatzell, M. Beidaghi, C.R. Dennison, E.C. Kumbur, Y. Gogotsi, "Activated carbon spheres as a flowable electrode in electrochemical flow capacitors", J. Electrochem. Soc. 161 (2014) A1078-A1083.